# Features of the Electronic and Charge States of Monovalent-Doped Manganite Films Probed by Magnetic Circular Dichroism

Yulia E. Samoshkina[a*], Dmitriy A. Petrov[a], Dmitry S. Neznakhin[b], Igor E. Korsakov[c], Andrei V. Telegin[d]

[a] *Kirensky Institute of Physics, Federal Research Center KSC SB RAS, 660036, Krasnoyarsk, Russian Federation*

[b] *Institute of Natural Sciences and Mathematics, Ural Federal University, 620002, Ekaterinburg, Russian Federation*

[c] *Moscow state University, 119991, Moscow, Russian Federation*

[d] *M.N. Mikheev Institute of Metal Physics UB of RAS, 620108, Ekaterinburg, Russian Federation*

[*] uliag@iph.krasn.ru

**Abstract**

Magnetic circular dichroism (MCD) spectroscopy in the range of 1.2 - 3.7 eV was studied for $La_{1-x}K_xMnO_3$ (x = 0.05 - 0.18) epitaxial films over a wide temperature range. The thin films were grown using a two-step procedure: deposition of $La_xMnO_{3-\delta}$ and potassium $K^+$ incorporation into the films via isopiestic annealing. The temperature behavior of the MCD effect in different spectral regions was analyzed alongside the temperature dependences of magnetization, resistivity, and magnetoresistance of the films. It was found that the MCD signal is sensitive not only to the magnetic but also to the charge sublattice of the material. Accordingly, a correlation between the magneto-optical and magnetoresistive responses of the system was identified. These findings underscore the high information content of MCD spectroscopy for investigating the magnetic and magnetotransport properties of strongly correlated magnetic oxides. The ground and excited electronic states in the $La_{1-x}K_xMnO_3$ films were identified, and the obtained data were compared with magneto-optical data for divalent - doped and lanthanum-deficient manganite films. Good agreement was observed, indicating the universality of the electronic structure and the shared mechanisms underlying the observed effects in such materials. These results broaden the understanding of the band structure in manganites and provide a solid foundation for its theoretical description.

## 1. Introduction

Complex manganese oxides constantly attract attention, demonstrating a variety of spin structures and combining various magnetic [1, 2], magneto-optical [3, 4], magnetocaloric [5, 6] and magnetotransport properties [7, 8]. Manganites based on $LnMnO_3$ (where Ln is a lanthanide), whose electron subsystem correlates with the magnetic subsystem, still remain one of the main objects of research [9, 10]. Due to their tendency to phase separation, they exhibit unique and practically significant properties, such as the metal-insulator transition [10], the colossal magnetoresistance effect [7], the giant magnetocaloric effect [5], and the giant magnetorefractive effect [3].

The primary focus is on divalent-doped manganites of $Ln_{1-x}R^{2+}_xMnO_3$, considering the distribution $Ln^{3+}_{1-x}R^{2+}_xMn^{3+}_{1-x}Mn^{4+}_xO_3$, as the colossal magnetoresistance effect near the phase transition point for $La_{0.7}Ca_{0.3}MnO_3$ films can reach 1300% [11]. It is noteworthy that monovalent-doped manganite films of $Ln_{1-x}R^+_xMnO_3$, taking into account the distribution $Ln^{3+}_{1-x}R^+_xMn^{3+}_{1-2x}Mn^{4+}_{2x}O_3$, demonstrate phase transitions (paramagnetic (PM) - ferromagnetic (FM) and metal-insulator) at higher temperatures and at lower $x$ values [12-18]. Moreover, such behavior has also been observed for lanthanum-deficient manganite films [13, 19, 20]. However, there are relatively few scientific studies of the materials mentioned, and even fewer in the form of thin films [e.g., 21-23]. In general, the relationship between the electronic and magnetic states in manganites remains unresolved, while clarification of this issue would allow predict their functional properties.

In this regard, the magneto-optical (MO) spectroscopy in the visible and near infrared spectral ranges is extremely useful as a direct source of information about the features of the electronic structure of materials. To obtain such information, the primary effects are magnetic circular dichroism (MCD) observed in transmitted light, along with a combination of ellipsometry and Kerr rotation (KR) observed in reflected light. Ultimately, the spectral dependences of the off-diagonal component of the permittivity tensor ($\varepsilon_{xy}$) are studied, directly reflecting the internal nature of the material. However, the MCD signal has an advantage over the KR signal, as it is directly proportional to the real part of $\varepsilon_{xy}$ ($\theta_{MCD} \sim \varepsilon'_{xy}$) [24]. Furthermore, the temperature behavior of the calculated off-diagonal component from KR spectroscopy has not been studied, which led to the ambiguity of the spectral analysis of $\varepsilon_{xy}$ in terms of electronic transitions. Therefore, the dependence of the KR spectrum shape on the conducting and insulating states of the material has not been given attention.

In addition, the field and temperature dependences of the MO signals reflect the magnetic behavior of the material, since both the MCD and KR effects are linear with magnetization. Thus, the intensity of these signals is expected to correlate with the magnetization of the material. However, the temperature evolution of the MCD signal observed in $Ln_{1-x}R^{2+}_xMnO_3$ films revealed differences in the spectral shape between the metallic and insulating phases of manganite within the same magnetic phase [25]. Previous studies have shown that the spectral shape of the MCD signal in $Ln_{1-x}R^{2+}_xMnO_3$ films is independent of film thickness, composition, or morphology, indicating the general nature of their electronic structure [25]. Moreover, a MO response typical of manganites undergoing a transition to the conducting state has been consistently observed, regardless of film thickness, morphology, or composition [26]. Notably, the high-sensitivity XANES (X-ray absorption near-edge structure) spectra exhibit essentially identical shapes for both the metallic and insulating phases of manganites [27].

This report presents the first investigation of visible-range MCD spectroscopy in $Ln_{1-x}R_xMnO_3$ manganites over a wide temperature range, with particular attention devoted to the magnetic and magnetotransport properties of the films.

## 2. Material and methods

Epitaxial films of $La_{1-x}K_xMnO_3$ (LKMO) with potassium doping 5%, 10%, 15%, and 18% were synthesized on a $SrTiO_3$ (STO) single-crystal substrate (001) with lattice parameter $c = 3.905$ Å. The films thickness was 100 nm. The samples were grown using the two-stage procedure: 1) $La_{1-x}MnO_{3-\delta}$ (LMO) epitaxial films were grown by the metal-organic chemical vapor deposition (MOCVD) method; 2) the $La_{1-x}MnO_{3-\delta}$ films were enriched with potassium by isopiestic annealing. The isopiestic annealing of $La_{1-x}MnO_{3-\delta}$ films with a powder mixture of $KNbO_3$ and $K_4Nb_6O_{17}$ providing a high concentration of volatile oxide in the gas phase leads to saturation of the samples with potassium due to gas phase transfer. An inert ceramic spacer separates the films and powder, preventing direct contact. The synthesis process using a two-step procedure is described in detail in [13, 28]. The annealing time of the studied samples was 8 hours in the flow of oxygen (0.21 atm) at T = 650 °C. The process of potassium incorporation into the LMO structure was monitored using energy-dispersive X-ray spectroscopy and by mass comparison with a ceramic target sample used as a standard.

The phase purity and crystal structure of the samples were examined by powder X-ray diffraction (XRD). XRD measurements were performed using CuKα radiation on a Bruker D8 ADVANCE diffractometer over the 20-80° range, with a step size of 0.02°. The XRD data (Fig. 1) showed that the samples consist of a highly oriented LKMO (001) phase without any traces of secondary phase. The lattice parameter ($c$) of the LKMO films with x varying from 0.05 to 0.18 is 0.3872, 0.3871, 0.3868, and 0.3863 nm, respectively. A decrease in lattice parameter for the LMO films is observed with increasing lanthanum deficiency, namely from 0.3876 to 0.3865 nm. The potassium incorporation into the LMO films as a result of annealing leads to a decrease in lattice parameter compared to the initial compositions. At the same time, the tendency for this parameter to decrease with increasing potassium content is preserved. The overall picture indicates elastic tension of the films on the substrate. For comparison, Figs. 1b and 1c show the dependence of the (003) reflex position on the $x$ for the LMO and LKMO films, respectively. Based on the lattice parameter values of the films, it follows that the samples before and after annealing show a pseudoperovskite cell. A similar value of *c* and XRD pattern were characteristic of $La_{0.88}MnO_{3-\delta}$ film grown on STO (001) substrate [20].

The MCD effect was measured using a specially developed automated system based on an MDR-12 monochromator and a Hamamatsu E678-11C photomultiplier. The MCD spectra were measured in normal geometry: both the magnetic field and the light beam were directed normal to the sample plane. The modulation of the polarization state of the light wave from the right-hand to the left-hand circular polarization relative to the magnetic field direction was used. The difference between the optical densities of the two waves obtained in this way was taken as the MCD signal. The spectral dependences of the MCD in the range of 1.2 - 3.5 eV were measured in the temperature range of 80 - 300 K and a magnetic field of 16 kOe. The MCD measurement technique is described in details [29].

The temperature and field dependences of the film magnetization were measured using the MPMS XL-7 EC SQUID magnetometer. The temperature interval was 3-300 K, and the magnetic field (H) was applied along to the sample plane and reached 10 kOe. The resistivity ($\rho$) of the films was measured in DC mode at a fixed current using the standard two-point probe technique, performed with a Keithley 2400 SourceMeter. Current was directed in the films plane. Temperature dependences were measured in the range of 90-300 K and H up to 8 kOe applied normally to the sample plane. The magnetoresistance was determined using the formula $\Delta\rho/\rho = (\rho_H - \rho_0)/\rho_0$, where $\rho_H$ and $\rho_0$ are the electrical resistance ($\rho$) value with and without an external magnetic field applied.

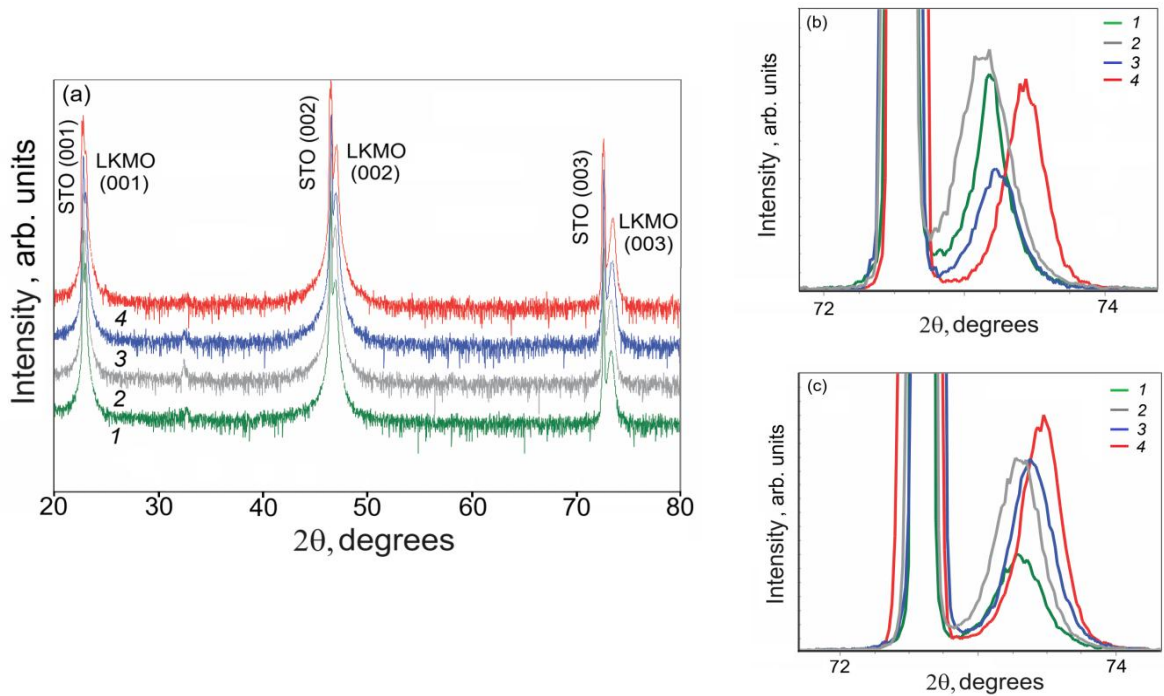

**Fig. 1.** a - X-ray diffraction patterns of the $La_{1-x}K_xMnO_3$ films. b - The dependence of the (003) reflex position on the index *x* for the $La_{1-x}MnO_{3-\delta}$ films. c - The dependence of the (003) reflex position on the potassium concentration for the $La_{1-x}K_xMnO_3$ samples. Curves 1-4 correspond to 0.05-0.18 of *x*, respectively.

## 3. Results

The temperature dependence of magnetization (*M*) shown in Fig. 2a indicates a phase transition to the ferromagnetic (FM) state in all LKMO films below room temperature. The Curie temperature ($T_C$) was defined as the temperature at which the extrapolation line intersects the abscissa (shown in Fig. 2a). Thus, the determined $T_C$ values are 150, 240, 260, and 280 K for x values of 0.05, 0.1, 0.15, and 0.18, respectively. The smooth change in magnetization near $T_C$, along with the absence of thermal hysteresis in the FC-FH curves measured under a magnetic field during cooling and heating, respectively, (inset in Fig. 2a) indicates a second-order magnetic phase transition [30].

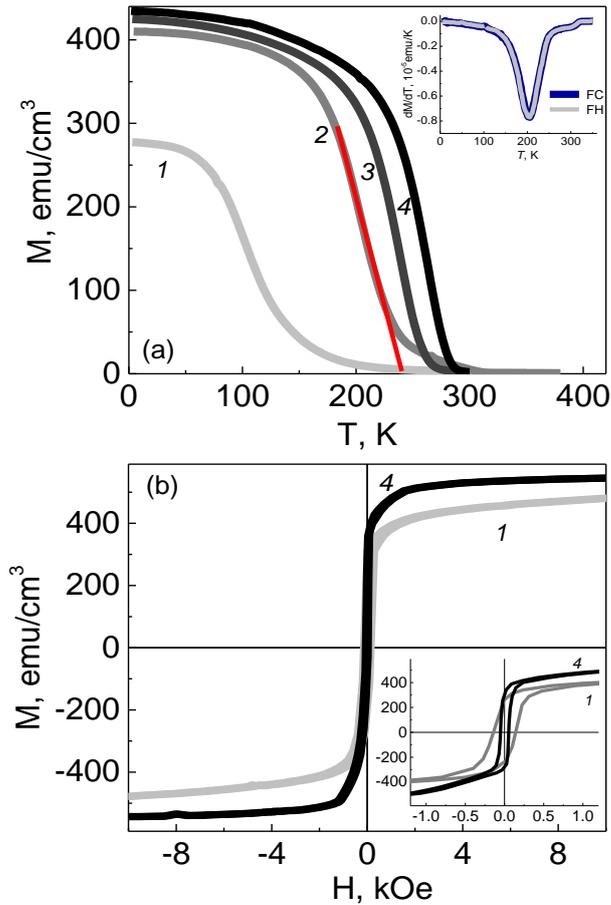

**Fig. 2.** Temperature (a) and magnetic field (b) dependences of magnetization (*M*) of the La$_{1-x}$K$_x$MnO$_3$ films taken at T = 3 K and H = 100 Oe. The magnetic field is applied along the film surface. The variation *x* of 0.05, 0.1, 0.15, and 0.18 is indicated by curves 1-4, respectively. Insets: a- dM/dT vs. T curves measured in FC-FH modes; b - Field dependences of *M* in low magnetic fields.

The $T_C$ value of the studied films is lower than that of polycrystals or ceramics with the same doping level [31, 32], which suggests a high defect density and/or the presence of strong epitaxial strains in the samples. However, a trend of increasing $T_C$ and the volume of the FM phase with increasing potassium content is also observed. The magnetization of the LKMO films increases and reaches saturation at low temperatures, which differs from the behavior of bulk polycrystalline samples of the same composition. The samples magnetization, in this case, after a sharp increase begins to decrease at low temperatures [12, 14, 16]. Such behavior was explained by the presence of the antiferromagnetic (AFM) regions within the FM phase. The contribution of these AFM inclusions is leveled out in high magnetic fields [16].

Low-temperature field dependences of the LKMO films demonstrate the low coercive force ($H_C$) of the samples (Fig. 2b). At the same time, $H_C$ decreases with increasing potassium content. Thus, $H_C$ is 140 Oe for x = 0.05, and $H_C$ is 50 Oe for x = 0.18 (inset in Fig. 2b). The clear rectangular shape of the loops indicates the in-plane anisotropy of the samples.

Normalized low-temperature MCD spectra of the LKMO films with different potassium contents are shown in Fig. 3. The spectral shape of the MCD signal for these films is described by three main absorption bands: two bands of the same sign in the region of 2 eV and 3.3 eV, as well as a band of the opposite sign near 2.6 eV. A shift of the bands to the low-energy region with increasing *x* has been observed. A similar shift in the bands position towards lower energies (beyond the error) is typical the hole-doped $Ln_{1-x}R^{2+}_xMnO_3$ manganites [33, 34]. This behavior is explained by the screening of the crystal field by the holes density, partially localized on the surrounding oxygen ions. The shape of the spectra is also similar to the MCD spectra of $La_{0.7}Sr_{0.3}MnO_3$, $La_{0.7}Ca_{0.3}MnO_3$ and $Pr_{1-x}Sr_xMnO_3$ films, studied previously [25, 26].

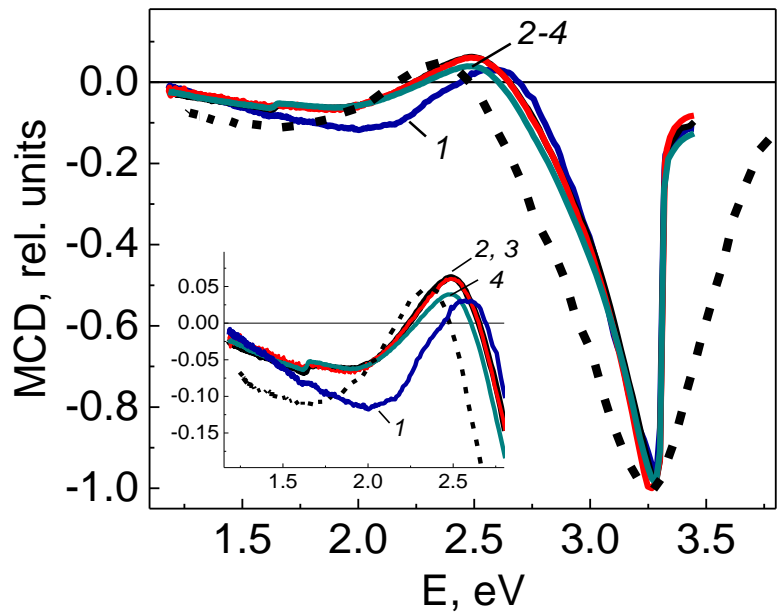

**Fig. 3.** Normalized MCD spectra of manganite films at T = 100 K: $La_{0.95}K_{0.05}MnO_3$ (curve 1), $La_{1-x}K_xMnO_3$ with *x* equal to 0.1, 0.15, and 0.18 (curves 2-4, respectively), and $La_{0.7}Ca_{0.3}MnO_3/LaAlO_3$ (dotted line). Inset: Enlarged view of the MCD spectra for the samples in the range of 1.2 - 2.8 eV.

For clarity, Fig. 3 shows the MCD spectrum of the epitaxial $La_{0.7}Ca_{0.3}MnO_3/LaAlO_3$ (LCMO) film [26]. The only notable difference for the LCMO sample is the shift of the MCD spectrum to the low-energy region by 0.1-0.2 eV and higher absorption edge energy. It should be noted that the most intense MCD band is located near the material's absorption edge, which for the LCMO film is 3.7 eV. In contrast, for the LKMO films, the absorption edge is 3.3 eV, meaning the intense MCD band is not fully visible. This is evident from the decomposition of the MCD spectra into components. Such decomposition was initially carried out for the $Ln_{1-x}R^{2+}_xMnO_3$ manganite films [34]. In this case, the MCD spectra were approximated using a minimum number of Gaussian lines. The fitting parameters were the line amplitude (A), position (E), and line width at half maximum ($\Delta E$). Similar to the results obtained for the $Ln_{1-x}R^{2+}_xMnO_3$ films [34], the spectra are well described by the $E_1$-$E_5$ lines at low temperatures (Fig. 4). The positions of all Gaussian lines are presented in Table 1, compared with the data for the LCMO film. Thus, the Gaussian lines $E_1$-$E_5$ correspond to the MCD bands *(1) - (5)*.

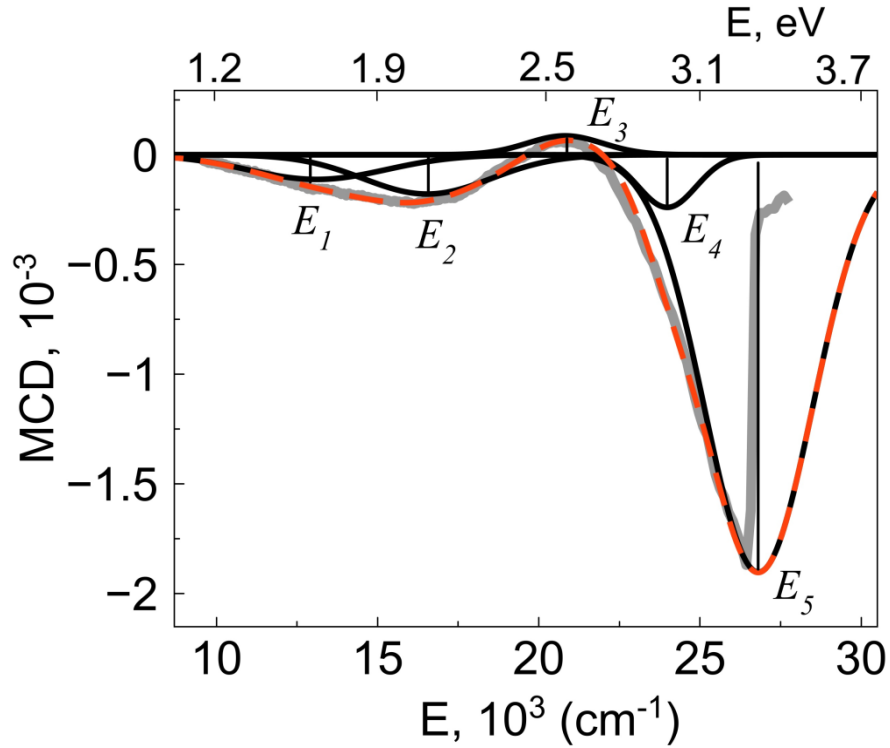

**Fig. 4.** Decomposition of the MCD spectrum (solid line) into Gaussian $E_1$ - $E_5$ lines for the $La_{0.95}K_{0.05}MnO_3$ film at T = 100 K. The sum of Gaussian lines is shown by the dotted line.

**Table 1.** The position (E) and nature of the Gaussian lines in the MCD spectra with decomposition at T = 100 K for the epitaxial LKMO and LCMO films.

| Composition | $E_1$ | | $E_2$ | | $E_3$ | | $E_4$ | | $E_5$ | |
|---|---|---|---|---|---|---|---|---|---|---|
| | cm$^{-1}$ | eV | cm$^{-1}$ | eV | cm$^{-1}$ | eV | cm$^{-1}$ | eV | cm$^{-1}$ | eV |
| LKMO (x=0.05) | 12809 | **1.59** | 16425 | **2.04** | 20772 | **2.58** | 24362 | **3.03** | 27133 | **3.37** |
| LKMO (x=0.10) | 12552 | **1.56** | 15862 | **1.97** | 20048 | **2.49** | 24106 | **2.99** | 26812 | **3.33** |
| LKMO (x=0.15) | 12483 | **1.55** | 15754 | **1.96** | 20050 | **2.49** | 23913 | **2.97** | 26651 | **3.31** |
| LKMO (x=0.18) | 12399 | **1.54** | 15700 | **1.95** | 20000 | **2.48** | 23832 | **2.96** | 26613 | **3.31** |
| LCMO (x=0.30) | 11514 | **1.43** | 14654 | **1.82** | 18970 | **2.36** | 23482 | **2.92** | 26573 | **3.3** |
| Nature | $^5E_g \rightarrow {}^5T_{2g}$ d–d transitions in Mn$^{3+}$ ions | | $^4A_{2g} \rightarrow {}^4T_{2g}$ d–d transitions in Mn$^{4+}$ ions | | | | $^4A_{2g} \rightarrow {}^4T_{1g}$ d–d transitions in Mn$^{4+}$ ions | | Mn$^{3+}$e$_g \rightarrow$ O$_{2p} \rightarrow$ Mn$^{4+}$e$_g$ charge transfer transitions | |

The MCD spectra of the LKMO films were measured at different temperatures, and the intensity variation of the bands *(1) – (5)* was traced. Data for the LKMO film with *x* = 0.05 are shown in Fig. 5. It was found that the intensity of the bands increases as the temperature decreases starting from T ≈ 205 K. The temperature dependence of all the band intensities is similar and follows the temperature behavior of the film magnetization (Fig. 5b). Notably, only the band *(5)* is visible in the spectrum above 210 K. Moreover, its intensity shows a decline at T = 205 K (inset in Fig. 5b) and increases again with rising temperature. This behavior in the MCD response of doped manganites is observed for the first time. The depicted minimum resembles the minimum typically seen in the temperature dependence of the FM resonance linewidth [35, 36]. Thus, it is likely associated with the transition from the pure FM phase to a paramagnetic (PM) region with FM inclusions. This high-temperature region may correspond to a Griffiths-like phase [37], which can be observed in the temperature range up to the maximum $T_C$ value of the LKMO system, which is 344 K [31].

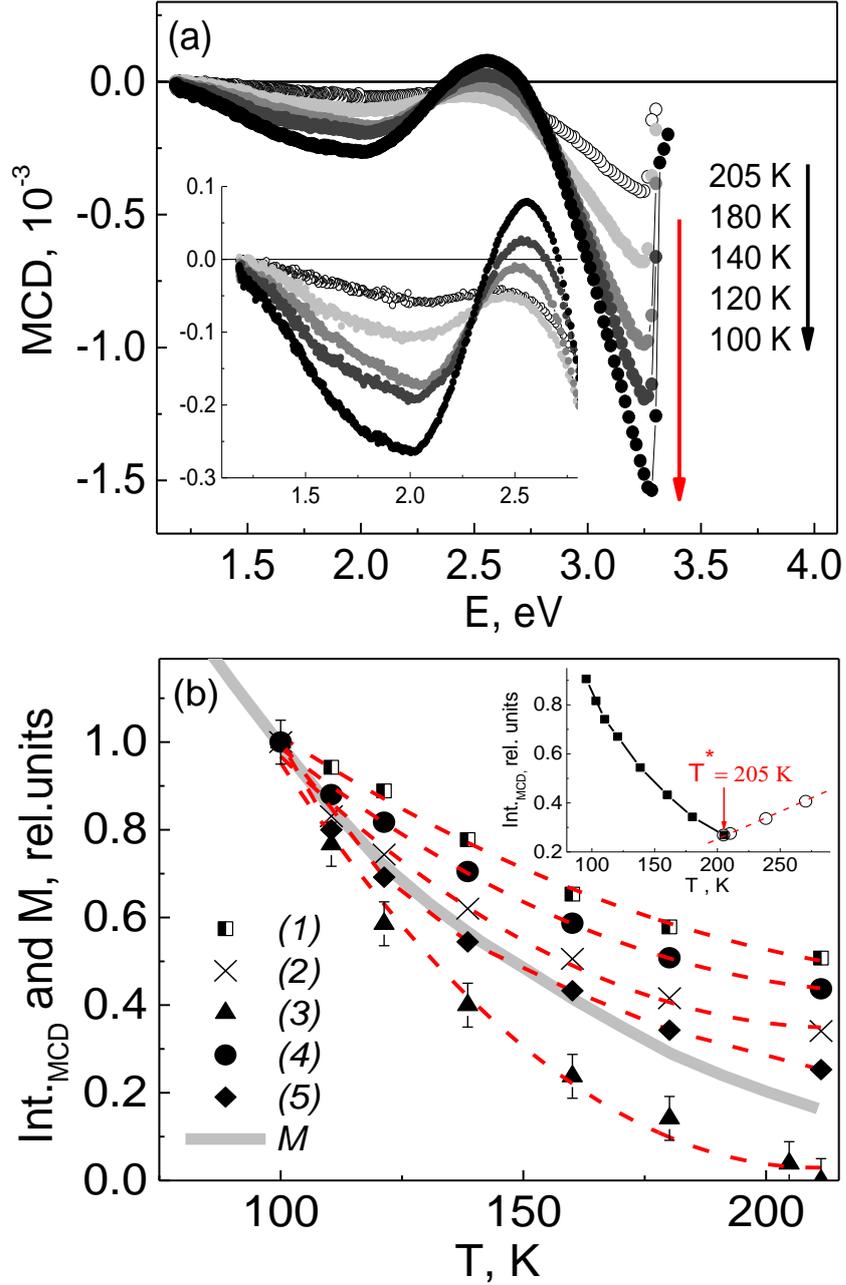

**Fig. 5** (a) Variation in the MCD spectra of the $La_{0.95}K_{0.05}MnO_3$ film as a function of temperature in the range of 100-205 K. Inset: Enlarged view of the MCD spectra for the sample in the range of 1.2 - 2.8 eV. (b) Normalized temperature dependences of the MCD band intensities (symbols) and the temperature dependence of the film magnetization (solid curve). The dotted lines are drawn to guide the reader's eye. Inset: General view of the normalized temperature dependence of the intensity of band (5) in the temperature range of 100-270 K. The magnetic field is 16 kOe for MCD spectra and 1 kOe for the magnetization measurements. The bars indicate the error in determining the band intensities.

In the case of the LKMO film with $x = 0.1$, the intensity of the observed MO bands increases as the temperature decreases over the entire studied range (Fig. 6). The intensity of bands *(1), (2), (4),* and

*(5)* follows the sample magnetization, while the temperature dependence of the intensity of band *(3)* differs from that of the magnetization (Fig. 6b). Notably, the MCD signal near 2.5 eV changes direction at T ≈ 235 K, and band *(3)* appears in the MCD spectrum as the temperature decreases further. A similar behavior has also been observed previously in polycrystalline $La_{0.7}Sr_{0.3}MnO_3$ and $Pr_{0.6}Sr_{0.4}MnO_3$ films near 2.3-2.4 eV, as well as the epitaxial LCMO film at 2.36 eV (Table 1) [26].

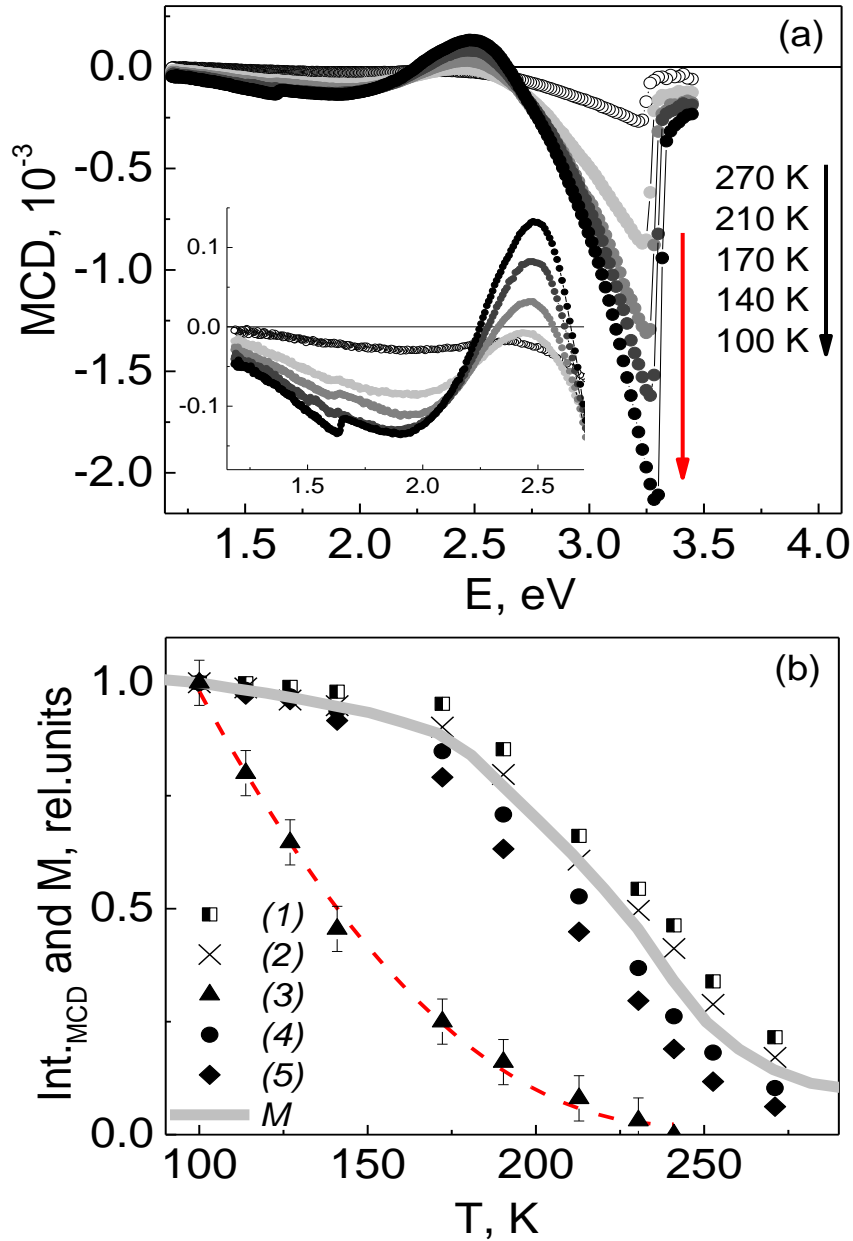

**Fig. 6** (a) Variation of the MCD spectra of the $La_{0.9}K_{0.1}MnO_3$ film as a function of temperature. Inset: Enlarged view of the MCD spectra for the samples in the range of 1.2 - 2.7 eV. (b) Normalized temperature dependences of the MCD band intensities (symbols) and the temperature dependence of the film magnetization (solid curve). The magnetic field is 16 kOe for MCD spectra and 1 kOe for magnetization measurements. The bars indicate the error in determining the band intensities. The dotted lines are drawn to guide the reader's eye.

For the films with $x = 0.15$ and $0.18$, a change in the behavior of several bands is observed (Fig. 7). Only the intensity of the high-energy band *(5)* follows the magnetization throughout the entire studied temperature range (Fig. 7b). The intensity of bands *(1)*, *(2)*, and *(4)* increases with decreasing temperature up to a certain point, first following the magnetization of the sample and then changing direction. The temperature behavior of the MCD signal and the band intensity near 2.5 eV is similar to that observed for the film with $x = 0.1$.

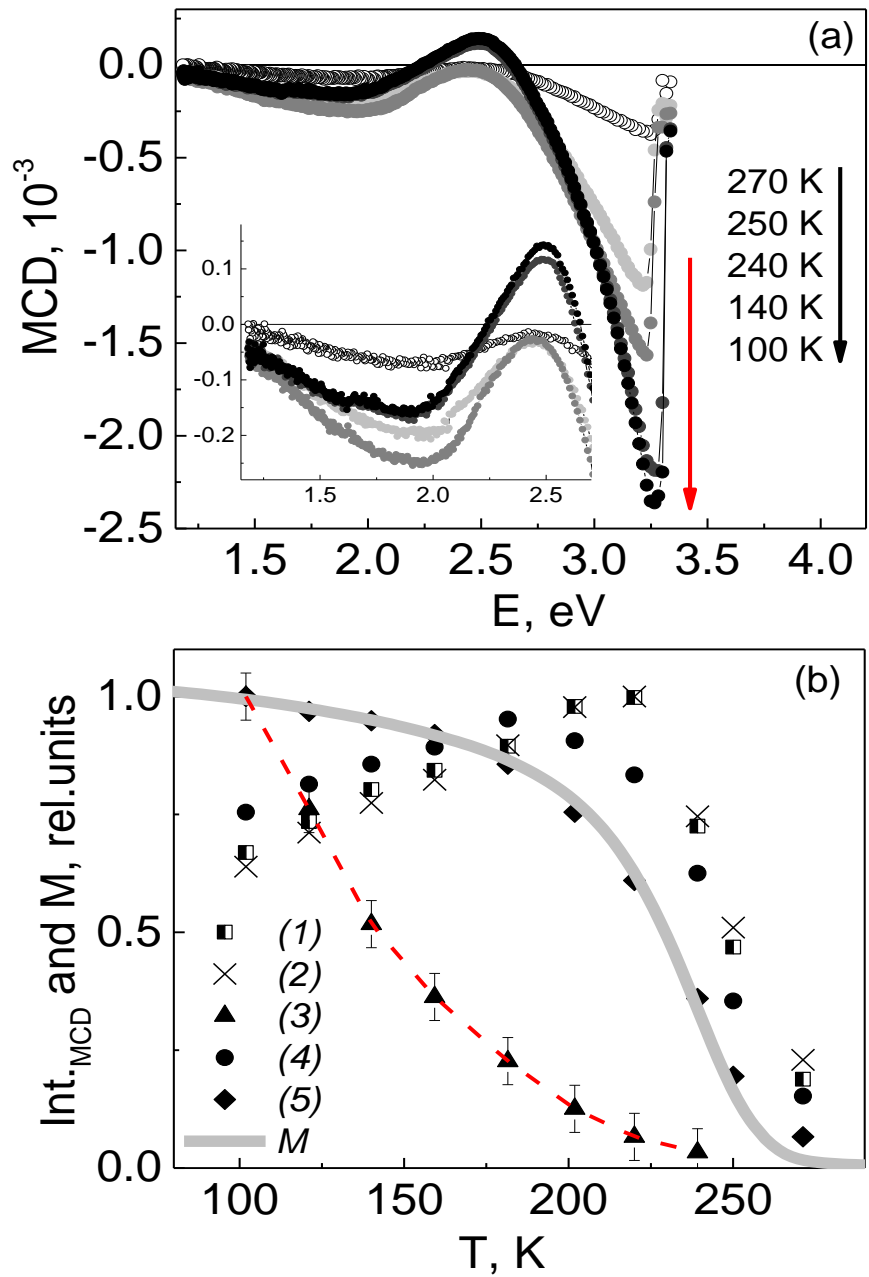

**Fig. 7** (a) Variation in the MCD spectra of the $La_{0.85}K_{0.15}MnO_3$ film as a function of temperature. Inset: Enlarged view of the MCD spectra for the samples in the range of 1.2 - 2.8 eV. (b) Normalized temperature dependences of the MCD bands intensity (symbols) and the temperature dependence of the film magnetization (solid curve). The magnetic field is 16 kOe for MCD spectra and 1 kOe for the magnetization measurements. The bars indicate the error in determining the band intensities. The dotted lines are drawn to guide the reader's eye.

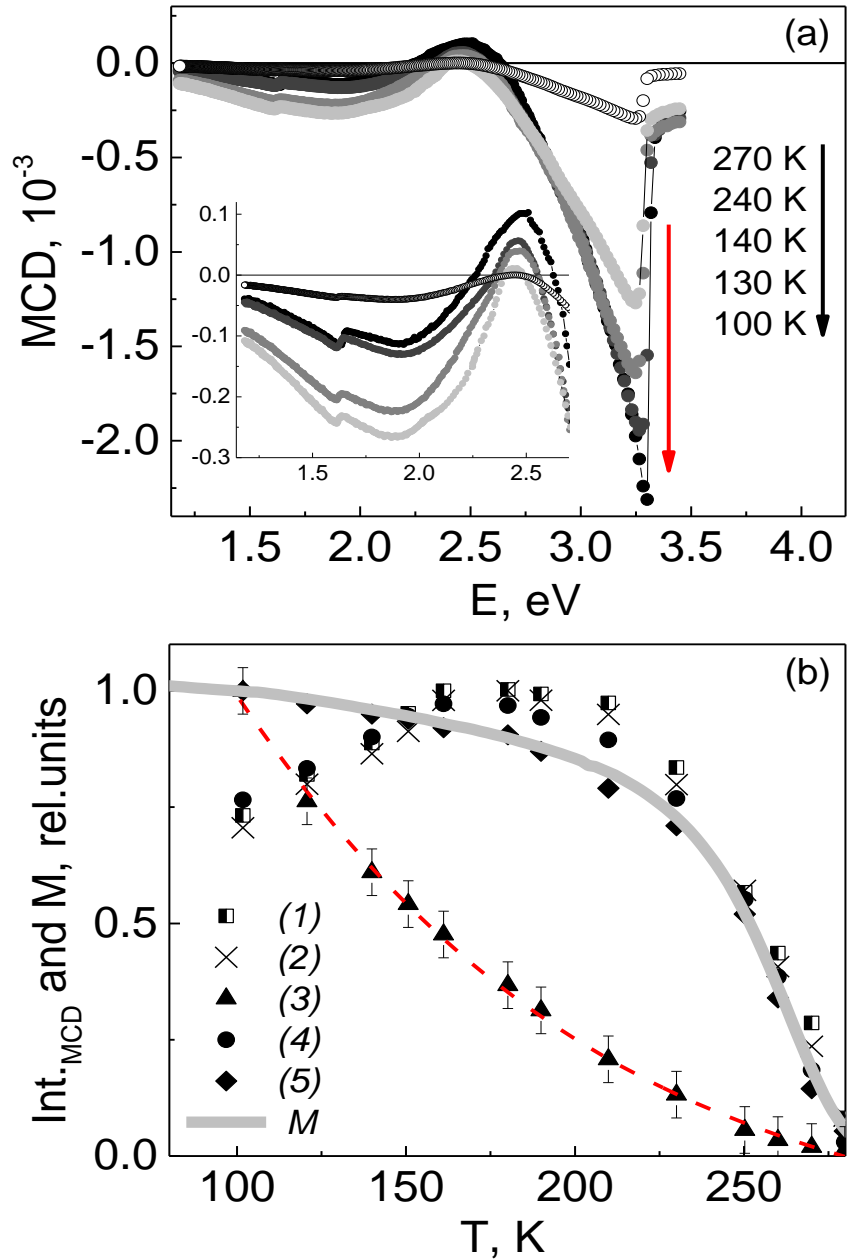

**Fig. 8** (a) Variation in the MCD spectra of the $La_{0.82}K_{0.18}MnO_3$ film as a function of temperature. Inset: Enlarged view of the MCD spectra for the samples in the range of 1.2 - 2.7 eV. (b) Normalized temperature dependences of the MCD bands intensity (symbols) and the temperature dependence of the film magnetization (solid curve). The magnetic field is 16 kOe for MCD spectra and 1 kOe for the magnetization measurements. The bars indicate the error in determining the band intensities. The dotted lines are drawn to guide the reader's eye.

## 4. Discussion

The temperature behavior of the intensity of the detected MCD bands coincides with that observed earlier for the $La_{0.7}Sr_{0.3}MnO_3$ and $Pr_{0.6}Sr_{0.4}MnO_3$ polycrystalline films of various thicknesses (20 - 150 nm) [25], as well as for 50 nm thick LCMO epitaxial film [26]. Therefore, it can be inferred that the nature of the observed MCD bands for LKMO is the same. In the $Ln_{1-x}R^{2+}_xMnO_3$ manganites, $Mn^{3+}$ and $Mn^{4+}$ ions are located in octahedral complexes of $(Mn^{3+}O_6)^{9-}$ and $(Mn^{4+}O_6)^{8-}$, respectively. Considering the effect of the crystal field solely within the octahedral complexes (the cluster model), the bands *(1)*, *(2)*, and *(4)* are associated with the spin-allowed d-d transition (Table 1). The energy of these bands corresponds well with the Tanabe-Sugano diagrams for the $d^3$ ($Mn^{4+}$) and $d^4$ ($Mn^{3+}$) electronic configurations [38, 39].

The presence of a kink in the temperature dependence of the intensity of bands *(1), (2),* and *(4)* (Figs. 7b and 8b) is likely associated with the differing growth rates of these bands compared to band *(3)*, which shows an opposite sign. Moreover, the observed phenomenon is influenced by the growth of the Drude term in the far infrared range as the temperature decreases, which is typical for metals [40]. This growth leads to a rearrangement of the optical conductivity and absorption spectra, as observed in [33, 41, 42] for lanthanum and neodymium, where the spectral weight shifts from higher energy (above 1 eV) to lower energy (below 1 eV). Therefore, the decrease in the intensity of the MO bands with decreasing temperature reflects the onset of the metallization process in manganites. In the case of the LKMO films with x = 0.05 and 0.1, the absence of a kink in the temperature dependence of the intensities of bands *(1), (2),* and *(4)* likely indicates an insufficient number of free charge carriers to initiate the metallization process.

The band associated with the transitions $Mn^{3+}e_{g1} \rightarrow O_{2p} \rightarrow Mn^{3+}e_{g2}$, often discussed in MO spectra [33, 43, 44], is likely formed below 1 eV. The most intense band *(5)* is obviously associated with the charge transfer transition $Mn^{3+}e_g \rightarrow O_{2p} \rightarrow Mn^{4+}e_g$ [33, 43, 44].

The appearance of the band *(3)* in the MCD spectra of the LKMO films is determined by the threshold temperature $T_S$ when the MCD signal changes direction. It has previously been established that a similar band is characteristic of the $Ln_{1-x}R^{2+}_xMnO_3$ manganites only in the conducting phase, i.e., it is observed at temperatures below the conventional metal-insulator transition ($T_{MI}$) [25, 26]. Therefore, there is a need the temperature dependences of the resistivity and magnetoresistance (MR) of the LKMO films to be additionally examined (Fig. 9). The values of conventional $T_{MI}$ (maximum point on the resistivity curve), temperature of maximum negative magnetoresistance ($T_{MR}$), and $T_S$ for the studied films are presented in Table 2. The MR value at $T_{MR}$ is 14, 8, 19, and 19% for films with x from 0.05 to 0.18, respectively. The data in the Table 2 indicate that the closeness of the $T_S$ and $T_{MR}$ values is characteristic of all LKMO films. Such a correlation could not be traced for polycrystalline manganite films, as the MR behavior of these samples in the low-temperature region is explained by the charge carrier scattering at the crystallite boundaries and tunnel magnetoresistance in the case of the textured films [26, 45]. However, the similar correlation between the $T_S$ and $T_{MR}$ values has been observed for the epitaxial LCMO film [26]. These data are presented in Table 2. It should be noted that the $T_{MI}$ value is proportional to the increase in the magnitude of the applied magnetic field. Therefore, in a magnetic field of 16 kOe, at which MCD was measured, the interval between the $T_S$ and $T_{MI}$ values will be even wider. Thus, the data for the epitaxial films suggest that band *(3)* represents the MO response of the system to a sharp change in conductivity under a magnetic field.

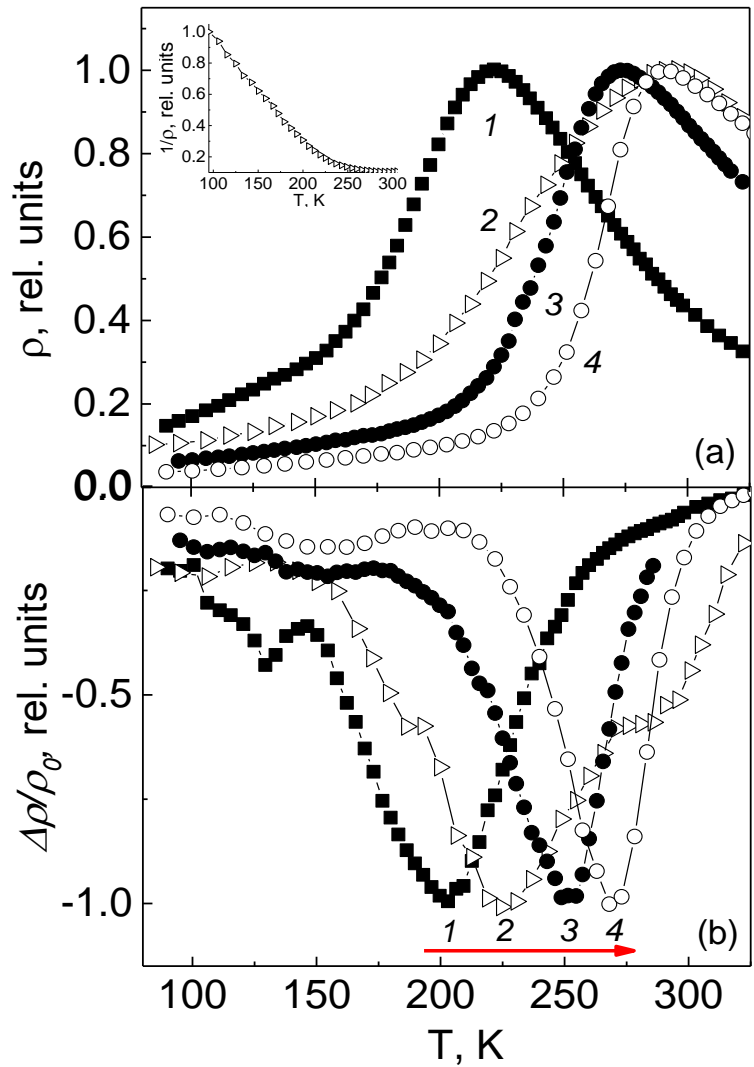

**Fig. 9**. Normalized temperature dependences of resistivity (a) and magnetoresistance (b) for the $La_{1-x}K_xMnO_3$ films with $x$ = 0.05, 0.1, 0.15, and 0.18 (curves 1-4, respectively). The magnetic field of 8 kOe is applied normal to the sample plane. Inset: Temperature dependence of the electrical conductivity of the $La_{0.9}K_{0.1}MnO_3$ film.

**Table 2.** The values of the metal-insulator transition temperature ($T_{MI}$), the temperature of maximum magnetoresistance ($T_{MR}$), the temperature at which the MCD signal at 2.5-2.6 eV changes its course ($T_S$), and the Curie temperature ($T_C$) for the LKMO films.

| Composition | $T_{MI}$, K | $T_{MR}$, K | $T_S$, K | $T_C$, K |
|---|---|---|---|---|
| LKMO | | | | |
| x = 0.05 | 220 | 205 | 205 | 150 |
| x = 0.10 | 290 | 230 | 235 | 240 |
| x = 0.15 | 270 | 250 | 240 | 260 |
| x = 0.18 | 290 | 270 | 270 | 280 |
| LCMO | 265 | 253 | 250 | 275 |

It should be noted that the temperature behavior of the band *(3)* intensity exhibits the same profile for all the studied LKMO films and is consistent with the profile of a similar band in $Ln_{1-x}R^{2+}_xMnO_3$ manganites [26]. For the $Ln_{1-x}R^{2+}_xMnO_3$ films, it has been previously shown that the temperature dependence of the band *(3)* intensity is maintained with changes in the film thickness and does not depend on the composition or morphology of the films. At the same time, it is similar to the electrical conductivity of the films (insert in Fig. 9a). Thus, for mono- and divalent-doped manganite films, the appearance of band *(3)* in the range from 2.3 to 2.6 eV reflects a change in the band structure of the material. Such behavior is typical for oxide semiconductors based on ZnO. When the zinc d-shell is completely filled, the MCD band near the bandgap edge of ZnO directly reflects different polarization states of charge carriers [46].

It is known that for bulk manganites and high-quality single-crystalline films the values of $T_C$, $T_{MI}$, and $T_{MR}$ are close [7]. The large discrepancy in these values for LKMO composition can be explained by the high density of defects in the films and/or the presence of epitaxial strain in them [25, 7]. This also explains the bend at the junction of the bands *(1)* and *(2)* (inset in Figs. 6a and 8a). For films with x above 0.05, such a bend is characteristic in the region of 1.6 eV and is clearly visible at temperatures below 200 K. As to the LKMO film with x = 0.05, this bend is hidden due to the difference in the magnitude of the MCD signal in the region of bands *(1)* and *(2)* (inset in Fig. 3). This is also indicated by the discrepancy in the temperature dependence of their intensity (Fig. 5b). A similar spectral bend was observed previously in the spectra of $Pr_{0.2}Sr_{0.8}MnO_3$ films, below the percolation threshold into a conducting high-temperature FM phase [34].

In addition, the LKMO film with x = 0.15 was further annealed in oxygen, and its magnetic and MO properties were examined. It was found that the film magnetization increased by 38% after annealing, while the $T_C$ value decreased by 35 K (Fig. 10a). At the same time, the coercivity remained unchanged (inset in Fig. 10a). This observed behavior suggests a reduction in defect density and/or relaxation of the epitaxial strain in the annealed film. More importantly, the spectral shape of the MCD signal remained unchanged. The same was true for the temperature dependences of the intensity of all the identified MCD bands (Fig. 10b). At the same time, the spectral bend in the region of 1.6 eV appeared only at a temperature of 120 K.

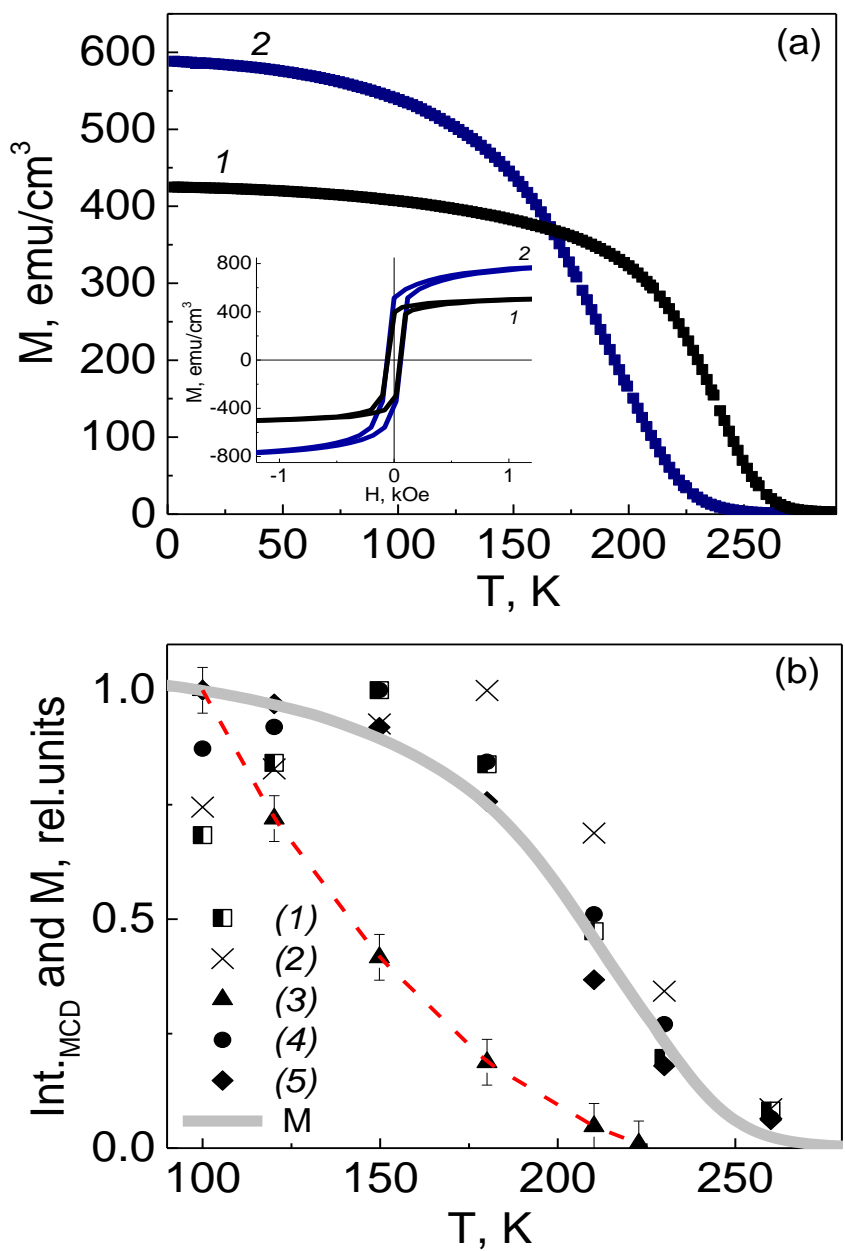

**Fig. 10** a - Field dependences of the $La_{0.85}K_{0.15}MnO_3$ films magnetization before (curve 1) and after (curve 2) annealing, T = 3 K. Inset: Temperature dependences of the same films at H = 100 Oe. The magnetic field is applied along the film surface. b - Normalized temperature dependences of the MCD bands intensity (symbols) and the temperature dependence of the $La_{0.85}K_{0.15}MnO_3$ film magnetization after annealing (solid curve). The magnetic field is 16 kOe for MCD spectra and 1 kOe for the magnetization measurements. The bars indicate the error in determining the band intensities.

The spectral shape of the MCD signal for the LKMO films is consistent with that of $Ln_{1-x}R^{2+}_xMnO_3$ films of varying morphology and thickness [25, 26]. Furthermore, the MCD spectra of the LKMO films closely resemble the KR spectra of $La_{1-x}Ag_xMnO_3$ films [15, 22]. Additionally, the MCD spectra of the specified materials are similar to the KR spectra of LMO samples [19]. It has been found that LMO films can exhibit a conducting FM state in the temperature range of 150-380 K, with $x$ varying from 0.02 to 0.34 [19, 20, 47, 48]. At the same time, the $T_C$ and $T_{MI}$ values of such films are sensitive to both the La/Mn ratio and the oxygen concentration. A similar behavior of the magnetization, resistivity, and MR between LMO [19, 20, 47, 48] and doped manganite films suggests the universality of the phase separation process, as well as the common mechanisms responsible for the observed effects. The optical spectra behavior of LMO films also resemble those of manganites with different types of doping [19, 40]. In particular, the metallization process in these films is accompanied by a transfer of spectral weight in optical density to the low-energy region as the temperature decreases [19, 40]. As noted earlier, this phenomenon is observed in the MO spectra of the LKMO films measured as a function of temperature (Figs. 7b and 8b). Therefore, the similarity in the spectral shapes of the optical and MO signals between the films of $La_{1-x}MnO_3$, $La_{1-x}R^+_xMnO_3$, and $Ln_{1-x}R^{2+}_xMnO_3$ confirms the unified electronic structure of these manganites in general. Moreover, the MO spectra shape of the manganite films under consideration is consistent with the KR spectra characteristic of $La_{0.85}Ba_{0.15}MnO_3$ and LSMO (x = 0.15-0.25) single crystals [49, 50]. This suggests that the observed features of the electronic structure presented in Table 1 are of a general nature and may be applicable across compositions of varying dimensionality.

## 5. Conclusions

It has been shown that the MCD spectra of $La_{1-x}K_xMnO_3$ (x = 0.05 - 0.18) films in the energy range of 1.2 - 3.7 eV exhibit three spin-allowed d-d electron transitions in $Mn^{3+}$ and $Mn^{4+}$ ions, as well as the charge-transfer transition of $Mn^{3+}e_g \rightarrow O_{2p} \rightarrow Mn^{4+}e_g$. The temperature behavior of these transitions correlates with that of the magnetic sublattice of the material. An MCD signal around 2.5 eV, reflecting the behavior of the charge sublattice, was also detected. It was found that the threshold temperature, $T_S$, at which this signal appears, closely matches the temperature of maximum magnetoresistance, $T_{MR}$. This suggests a correlation between the magneto-optical and magnetoresistive behaviors of the system. Additionally, the temperature profile of the MCD bands intensity associated with d-d transitions reflects the metallization process in the samples. In this case, the bands intensity decreases with decreasing temperature. The increase in the intensity of the $Mn^{3+}e_g \rightarrow O_{2p} \rightarrow Mn^{4+}e_g$ band with increasing temperature indicates a Griffiths-like phase. Thus, MCD spectroscopy offers valuable insights into the electronic, charge, and magnetic states of the manganite films.

Good agreement between the shape of MCD spectra and their consistent changes with temperature has been observed in the manganite films with both mono- and divalent doping. This suggests that the mechanisms responsible for the magnetic and magnetotransport properties are shared across these strongly-correlated materials. Additionally, the similarity in the low-temperature behavior of optical and magneto-optical spectra in doped and lanthanum-deficient manganites indicates the universality of the electronic structure, as well as the common mechanisms responsible for the observed effects in such materials. These results provide a solid basis for the theoretical description of the band structure of multifunctional manganite-based materials. Such an approach will bring researchers closer to understanding the nature of the observed effects and to controlling the practically significant properties of magnetic oxides.

**CRediT authorship contribution statement**

**Yulia E. Samoshkina**: Conceptualization, Methodology, Formal analysis, Visualization, Writing – original draft, Writing – review & editing. **Dmitriy A. Petrov**: Investigation, Validation. **Dmitry S. Neznakhin:** Resources, Investigation, Validation. **Igor E. Korsakov**: Methodology, Resources, Investigation, Formal analysis. **Andrei V. Telegin**: Investigation, Validation, Formal analysis, Writing – review & editing.

**Declaration of Competing Interest**

The authors declare that they have no known competing financial interests or personal relationships that could have appeared to influence the work reported in this paper.

**Acknowledgments**
The work was carried out within the state assignment of Kirensky Institute of Physics. The contribution of I.E. Korsakov was supported by the state assignment of Moscow state University. The contribution of A.V. Telegin was supported within the framework of the state assignment of the Ministry of Science and Higher Education of the Russian Federation for the IMP UB RAS.

**Data Availability**

Data will be made available on request.